\documentclass[preprint,preprintnumbers,a4paper,nofootinbib]{revtex4-1}

\usepackage{amsmath,amssymb}

\usepackage[utf8]{inputenc}
\usepackage{graphicx, color}

\begin{document}

\preprint{UTHEP-740, UTCCS-P-126, KANAZAWA-19-08, NCTS-CMR/1902}

\begin{center}
  {\Large{\textbf{Investigation of complex $\phi^{4}$ theory at finite density\\ in two dimensions using TRG}}}
  \\
  \medskip
  \vspace{1cm}
  \textbf{
    Daisuke~Kadoh$^{\,a,b,}$\footnote{kadoh@keio.jp},
    Yoshinobu~Kuramashi$^{c,}$\footnote{kuramasi@het.ph.tsukuba.ac.jp},
    Yoshifumi~Nakamura$^{d,}$\footnote{nakamura@riken.jp},
    Ryo~Sakai$^{e,}$\footnote{ryo-sakai@uiowa.edu},
    Shinji~Takeda$^{f,}$\footnote{takeda@hep.s.kanazawa-u.ac.jp},
    Yusuke~Yoshimura$^{c,}$\footnote{yoshimur@ccs.tsukuba.ac.jp}
  }
  \bigskip

  $^a$ {\small Physics Division, National Center for Theoretical Sciences, National Tsing-Hua University, Hsinchu, 30013, Taiwan}

  $^b$ {\small Research and Educational Center for Natural Sciences, Keio University, \\ Yokohama 223-8521, Japan}

  $^c$ {\small Center for Computational Sciences, University of Tsukuba, Tsukuba 305-8577, Japan}

  $^d$ {\small RIKEN Center for Computational Science, Kobe 650-0047, Japan}

  $^e$  {\small Department of Physics and Astronomy, The University of Iowa, Iowa City, IA 52242, USA}

  $^f$ {\small Institute for Theoretical Physics, Kanazawa University, Kanazawa 920-1192, Japan}

\end{center}

\vspace{0cm}







\date{\today \\
  \vspace{5mm}
}

\begin{abstract}
  We study the two-dimensional complex $\phi^{4}$ theory at finite chemical potential
  using the tensor renormalization group.
  This model exhibits the Silver Blaze phenomenon in which bulk observables are independent of
  the chemical potential below the critical point.
  Since it is expected to be a direct outcome of an imaginary part of the action,
  an approach free from the sign problem is needed.
  We study this model systematically changing the  chemical potential in order to check the applicability of the tensor renormalization group to the model in which scalar fields are discretized by the Gaussian quadrature.
  The Silver Blaze phenomenon is successfully confirmed on the extremely large volume $V=1024^2$ and
  the results are also ensured by another tensor network representation with a character expansion.
\end{abstract}

\maketitle

\section{Introduction}
\label{sec:introduction}

The tensor network (TN) is a promising approach to study lattice models
with a sign problem.
Coarse-graining algorithms of tensor networks such as the tensor renormalization group (TRG) \cite{Levin:2006jai}
do not have any stochastic process unlike the Monte Carlo method which is based
on the stochastic interpretation of the Boltzmann factor in path integrals.
So a development of this approach could lead to deep understanding of quantum field theories that suffer from the sign problem
such as  QCD at finite chemical potential, finite $\theta$ angle,
chiral gauge theories and SUSY theories.
Although the TRG algorithm has been already introduced into the research of
lattice quantum field theories~\cite{Shimizu:2012wfa,
  Liu:2013nsa,Yu:2013sbi,Denbleyker:2013bea,Shimizu:2014uva,Unmuth-Yockey:2014iga,
  Shimizu:2014fsa,Takeda:2014vwa,Kawauchi:2016xng,Meurice:2016mkb, Sakai:2017jwp,
  Yoshimura:2017jpk,Shimizu:2017onf,Kadoh:2018hqq,Kuramashi:2018mmi, Kadoh:2018tis,
  Kuramashi:2019cgs},
further studies are desirable to confirm
if the TRG properly works for theories with a severe sign problem.

The complex $\phi^{4}$ theory at finite chemical potential is the simplest model that
suffers from a severe sign problem.
This model exhibits the so-called Silver Blaze phenomenon
in which bulk observables
do not depend on the chemical potential below the critical point.
Since it is directly related to the imaginary part of the action,
various methods that could overcome the sign problem, such as
the complex Langevin approach \cite{Aarts:2008wh}, the thimble method \cite{Cristoforetti:2013wha,Fujii:2013sra,Mori:2017nwj},
and the worldline representation \cite{Gattringer:2012df,Orasch:2017niz},  have been used to study the model.
In case of TRG, it is not straightforward to apply the algorithm to the
scalar field theory because the tensor indices are given by the field variable which takes any real or complex number
and numerical computation is not directly applied to such an infinite dimensional tensor.

In refs.~\cite{Kadoh:2018hqq,Kadoh:2018tis} we have proposed a methodology of defining a finite dimensional tensor
in the scalar field theory. We employed the
Gaussian quadrature rule to discretize the scalar field so that  a critical coupling constant
of the $Z_2$ symmetry breaking in the two-dimensional real $\phi^{4}$ theory is evaluated with the TRG procedure.
The result was consistent with those
obtained with other conventional methods. Namely, our discretization method effectively works for the real scalar field theory.
This implies that the TRG approach with the discretized field variables can be also effective for a complex scalar field theory.

In this paper, we study the  two-dimensional complex $\phi^{4}$ theory at finite chemical potential
using the TRG method with the Gauss quadrature discretization for the scalar field.
The expectation values for the scalar field and the number density are evaluated to investigate
the Silver Blaze phenomenon.
Furthermore, in order to confirm that the
TRG method properly works,
we compare the results to those obtained from another TN representation with the character expansion.

The rest of this paper is organized as follows:
In Sec.~\ref{sec:2dcphi4} we define the target model and construct the TN representation for the partition function. Numerical results are presented in Sec.~\ref{sec:nresults}, where the Silver Blaze phenomenon
is confirmed. We also make a comparison of the results obtained from the naive TN representation
of the partition function and another TN representation. Section~\ref{sec:summary} is devoted to summary and future perspectives.

\section{Two-dimensional complex $\phi^{4}$ theory}
\label{sec:2dcphi4}

The Euclidean continuum action of the two-dimensional complex $\phi^{4}$
theory at finite chemical potential is defined by
\begin{align}
  \label{eq:caction}
  S_{\mathrm{cont}} = \int \mathrm{d}^2 x
  \left\{
  \sum_{\nu=1}^2\left| \partial_{\nu} \phi \right|^{2}
  + \left( m^{2} - \mu^{2} \right) \left| \phi \right|^{2}
  + \mu \left( \phi^{*} \partial_{2}\phi - \phi \partial_{2}\phi^{*} \right)
  + \lambda \left| \phi \right|^{4}
  \right\}
\end{align}
with a complex scalar field $\phi(x)$, the bare mass $m$, the quartic coupling constant $\lambda > 0$,
and the chemical potential $\mu$.
This theory describes a relativistic Bose gas with finite chemical potential.
The action is complex for $\mu \neq 0$ because the third term of
eq.~(\ref{eq:caction}) is a pure imaginary number.

In the lattice theory,  the scalar field denoted as $\phi_n$ lives on a site $n$
of a lattice $\Gamma=\{ (n_1,n_2) \, | \, n_\nu=1,2,\ldots, N_i \, \}$ with the lattice volume $V=N_1\times N_2$.
The lattice spacing $a$ is set to $1$.
We assume that the scalar field satisfies the periodic boundary condition,
$\phi_{n + N_\nu \hat \nu}=\phi_n$ for $\nu=1,2$, where $\hat \nu$ is the unit vector of the $\nu$-direction.
The lattice action is given by
\begin{align}
  \label{eq:laction}
  S = \sum_{n\in\Gamma}
  \left[
  \left( 4 + m^{2} \right) \left| \phi_{n}^{2} \right|
  + \lambda \left| \phi_{n} \right|^{4}
  - \sum_{\nu=1}^{2} \left( e^{\mu \delta_{\nu 2}} \phi_{n}^{*}\phi_{n+\hat{\nu}} + e^{-\mu \delta_{\nu 2}} \phi_{n}\phi_{n+\hat{\nu}}^{*} \right)
  \right].
\end{align}
Note that the chemical potential is introduced as a pure imaginary constant vector potential
in the temporal direction \cite{Hasenfratz:1983ba}.
Since the lattice action also satisfies $(S(\mu))^*=S(-\mu)$, it is difficult to apply a naive Monte Carlo method to this model.

The partition function is defined as a standard manner:
\begin{align}
  \label{eq:pfunction}
  Z = \int \mathcal{D} \phi \, e^{-S},
\end{align}
where the complex field $\phi_n$ is represented in terms of two real fields as
$\phi_{n} = \frac{1}{\sqrt{2}}\left( A_{n} + i B_{n} \right)$
and
the integral measure is given by
$\mathcal{D} \phi   \equiv \prod_{n\in\Gamma} \mathrm{d} A_{n} \mathrm{d} B_{n}$.
In the following we show that $Z$ is represented as a tensor network
according to refs.~\cite{Kadoh:2018hqq,Kadoh:2018tis}.
The expectation value of any local field can also be represented as a tensor network in a similar way.

The Boltzmann weight $e^{-S}$ is expressed as a product of local factors:
\begin{align}
  \label{eq:bweight}
  e^{-S} = \prod_{n\in\Gamma} f_{1}\left( \phi_{n}, \phi_{n+\hat{1}} \right)   f_{2}\left( \phi_{n}, \phi_{n+\hat{2}} \right),
\end{align}
where
\begin{align}
  \label{eq:lfactor}
  f_{\nu} (w, z)
  = \exp
  \left\{
  - \left( 1+\frac{m^{2}}{4} \right) \left( \left| w \right|^{2} + \left| z \right|^{2} \right)
  - \frac{\lambda}{4} \left( \left| w \right|^{4} + \left| z \right|^{4} \right)
  + e^{\mu \delta_{\nu 2}} w^{*}z + e^{-\mu \delta_{\nu 2}} w z^{*}
  \right\}
\end{align}
for $w, z \in \mathbb{C}$. It is possible to decompose the  Boltzmann weight in this way as long as
the lattice action contains only the nearest-neighbor interaction.

The continuous scalar field  is discretized by the Gauss--Hermite quadrature rule
to introduce a finite dimensional tensor as in refs.~\cite{Kadoh:2018hqq,Kadoh:2018tis}.
For one-variable integration of a proper function $g(x)$,
the quadrature provides a discretization as follows:
\begin{align}
  \label{eq:GHquadrature}
  \int_{-\infty}^{\infty} \mathrm{d}x \, e^{-x^{2}} g\left( x \right)
  \approx \sum_{\alpha=1}^{K} w_{\alpha} g\left( y_{\alpha} \right)
\end{align}
where $y_{\alpha}$ and $w_{\alpha}$ are the $\alpha$-th root of the $K$-th Hermite polynomial $H_K(x)$
and the corresponding weight defined as $w_\alpha=2^{K-1}K!\sqrt{\pi}/(K^2H_{K-1}(y_\alpha)^2)$, respectively.
Here $K$ dictates the order of approximation and for large $K$ the accuracy of approximation is expected to be better
\footnote{This depends on $g(x)$.
  In an actual computation, we check the convergence of result by increasing $K$.
}.

For the two-variable case ($\phi=\frac{1}{\sqrt{2}}(A+i B)$ with $A, B\in\mathbb{R}$),
we have
\begin{align}
  \label{eq:GHquadrature_2}
  \int_{-\infty}^{\infty} \mathrm{d} A
  \int_{-\infty}^{\infty}
  \mathrm{d} B \,\,e^{-2|\phi|^2} h\left( \phi \right)
  \approx \sum_{\alpha=1}^{K} \sum_{\beta=1}^{K} w_{\alpha}w_{\beta}
  \, h( \phi(\alpha,\beta)),
\end{align}
where
\begin{align}
  \label{complex_field_from_ab}
  \phi(\alpha,\beta) \equiv  \frac{y_{\alpha}+iy_{\beta}}{\sqrt{2}}.
\end{align}
Applying eq.~(\ref{eq:GHquadrature_2}) to each complex field,
$Z$ is approximated by $Z(K)$ as
\begin{align}
  \label{eq:dpfunction}
  Z\approx
  Z\left( K \right) = \sum_{\left\{ \alpha, \beta \right\}}
  \prod_{n\in\Gamma} w_{\alpha_{n}} w_{\beta_{n}} \exp\left( y_{\alpha_{n}}^{2} + y_{\beta_{n}}^{2} \right)
  \prod_{\nu=1}^{2} f_{\nu}\left( \phi(\alpha_{n}, \beta_{n}), \phi( \alpha_{n+\hat{\nu}}, \beta_{n+\hat{\nu}}) \right),
\end{align}
where
$\sum_{\left\{ \alpha, \beta \right\}}
\equiv
\prod_{n\in\Gamma} \sum_{\alpha_{n}=1}^{K} \sum_{\beta_{n}=1}^{K}$.

As a result of the discretization, $f_\nu$ can be regarded as a $K^{2} \times K^{2}$ complex valued matrix:
\begin{align}
  \label{eq:matrix_f}
  M^{[\nu]}_{\alpha\beta,\alpha^\prime\beta^\prime}
  \equiv
  f_{\nu} ( \phi(\alpha, \beta), \phi(\alpha^\prime, \beta^\prime) )
\end{align}
with the row index $\alpha, \beta=1,2,\ldots,K$ and the column index $\alpha', \beta'=1,2,\ldots,K$.
Note that $\phi(\alpha,\beta)$ is given by discretized points $y_\alpha, y_\beta$ in eq.~(\ref{complex_field_from_ab}).
Then the singular value decomposition is applied to the matrix:
\begin{align}
  \label{eq:svd_f}
  M^{[\nu]}_{\alpha\beta, \alpha^\prime\beta^\prime}
  = \sum_{k=1}^{K^{2}} U^{[\nu]}_{\alpha\beta, k} \sigma^{[\nu]}_{k} V^{[\nu] \dagger}_{k, \alpha^{\prime}\beta^{\prime} },
\end{align}
where $\sigma^{[\nu]}_{k}$ is $k$-th singular value sorted in the descending order,
and $U^{[\nu]}$ and $V^{[\nu]}$ are $K^2 \times K^2$ unitary matrices
with the row index $\alpha, \beta$ and the column index $k$.
Plugging eq.~(\ref{eq:svd_f}) into eq.~(\ref{eq:dpfunction}),
we find that  $Z(K)$ can be expressed as a tensor network,
\begin{align}
  \label{eq:tnrep_z}
  Z\left( K \right) = \sum_{\left\{ x, t \right\}} \prod_{n\in\Gamma} T_{x_{n} t_{n} x_{n-\hat{1}} t_{n-\hat{2}}},
\end{align}
where
\begin{align}
  \label{eq:tensor}
  T_{ijkl}
  = \sqrt{\sigma^{[1]}_{i} \sigma^{[2]}_{j} \sigma^{[1]}_{k} \sigma^{[2]}_{l}} \sum_{\alpha, \beta = 1}^{K} w_{\alpha} w_{\beta}
  \exp\left( y_{\alpha}^{2} + y_{\beta}^{2} \right)
  U^{[1]}_{\alpha \beta, i} U^{[2]}_{\alpha \beta, j} V^{[1] \dagger}_{k,\alpha \beta} V^{[2] \dagger}_{l, \alpha \beta}
\end{align}
and $\sum_{\left\{ x, t \right\}} \equiv \prod_{n\in\Gamma} \sum_{x_{n},t_{n}=1}^{K^{2}}$.

We obtain a finial expression  by truncating the summation in eq.~(\ref{eq:tnrep_z}) up to $D$ $(\leq K^{2})$ to reduce the computational complexity:
\begin{align}
  \label{eq:rtnrep_z}
  Z\left( K \right) \approx
  {\sum_{\left\{ x, t \right\}}}^\prime  \prod_{n\in\Gamma}
  T_{x_{n} t_{n} x_{n-\hat{1}} t_{n-\hat{2}}},
\end{align}
where  $\sum^\prime_{\left\{ x, t \right\}} \equiv \prod_{n\in\Gamma} \sum_{x_{n},t_{n}=1}^{D}$.
This truncation keeps a better precision when $\sigma_k^{[\nu]}$ in eq.~(\ref{eq:svd_f}) has a sharp hierarchy structure.
We should note that the initial tensor $T$ depends on $K$.
$D$ becomes the bond dimension of tensors which is fixed throughout computations,
and the convergence of results for $K$ and $D$ are checked numerically.

\section{Numerical results}
\label{sec:nresults}

Numerical results of two-dimensional complex $\phi^4$ theory
at finite chemical potential are presented in this section.
The TRG~\cite{Levin:2006jai} is employed to coarse-grain the tensor network eq.~(\ref{eq:rtnrep_z})
on a periodic lattice with the volume $V=N^2$ ($N=2^m, m \in \mathbb{Z}$) and the lattice spacing $a=1$.
The coarse-graining procedure of partition function is briefly described in our previous paper~\cite{Kadoh:2018tis}
in which a procedure for the expectation value of a local field is also given.
In the TRG algorithm, the SVD is truncated up to a fixed integer $D$, which is the bond dimension of tensors.

\begin{figure}[htbp]
  \centering
  \includegraphics[width=0.8\hsize]{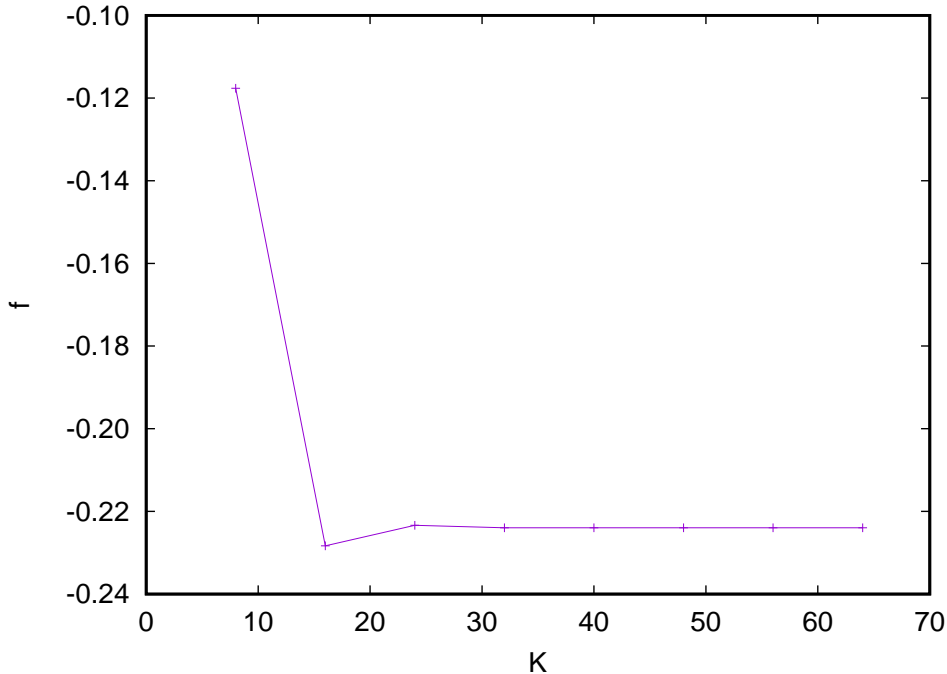}
  \caption{Free energy density for $D=64$ and $m^{2}=0.01$, $\lambda=\mu=1$ on $V=1024^2$.}
  \label{fig:K_vs_F}
\end{figure}
\begin{figure}[htbp]
  \centering
  \includegraphics[width=0.8\hsize]{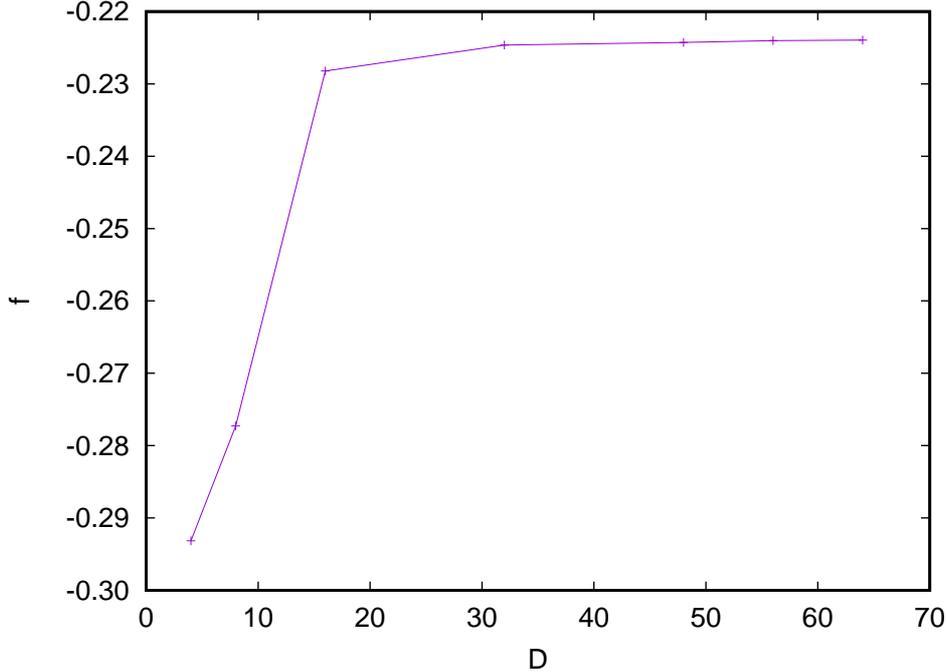}
  \caption{Free energy density for $K=64$ and $m^{2}=0.01$, $\lambda=\mu=1$ on $V=1024^2$.}
  \label{fig:D_vs_F}
\end{figure}
Figures~\ref{fig:K_vs_F} and~\ref{fig:D_vs_F} show the $K$-dependence and the $D$-dependence
of the free energy density $f=-\frac{1}{V} {\rm ln} Z$
for a typical parameter set.
$K$ is the number of points used  in the discretization of scalar fields
as presented in eq.~(\ref{eq:GHquadrature_2}). The initial tensor eq.~(\ref{eq:tensor}) depends on $K$
since it is made of $K$-dependent unitary matrices
associated with $M$ in eq.~(\ref{eq:matrix_f}).
The result converges as both $K$ and $D$ increase, and $K=D=64$, which are fixed in the following,
are large enough to obtain converged results.

\subsection{Average phase factor}
\label{sec:aphase}

Let  $ \langle \cdot \rangle_{\rm pq}$ be
an expectation value in the phase quenched theory with
partition function,
\begin{align}
  \label{eq:zpq}
  Z_{\mathrm{pq}} = \int \mathcal{D} \phi \, e^{-\mathrm{Re}\left( S \right)}.
\end{align}
Then the expectation value of an operator ${\cal O}$ may be expressed as
\begin{align}
  \label{ratio_full}
  \langle {\cal O}\rangle
  =
  \frac{\langle {\cal O}e^{i\theta}\rangle_{\rm pq}}{\langle e^{i\theta}\rangle_{\rm pq}},
\end{align}
where $e^{-S}=e^{-\mathrm{Re}\left( S \right)}e^{i\theta}$.
Using the TRG, $Z_{\rm pq}$ and $\langle {\cal O}\rangle_{\rm pq}$ for a local operator $ {\cal O}$ can also be evaluated
from a tensor dropping the last two terms in eq.~(\ref{eq:lfactor}).

The sign problem appears as a difficulty in evaluating the ratio of eq.~(\ref{ratio_full}).
For large $\mu$,
since the phase factor $e^{i\theta}$
has a large fluctuation,
both the average phase factor,
\begin{align}
  \langle e^{i\theta}\rangle_{\rm pq}
  =\frac{Z}{Z_{\mathrm{pq}}},
  \label{average_phase}
\end{align}
and $\langle {\cal O}  e^{i\theta} \rangle_{\rm pq}$ approach zero.
Then,  in the Monte Carlo method, it becomes difficult to evaluate   $\langle {\cal O}\rangle$ due to a $0/0$ problem.
In other words, the  severeness of the sign problem is measured by the numerical value of eq.~(\ref{average_phase}).

\begin{figure}[htbp]
  \centering
  \includegraphics[width=0.8\hsize]{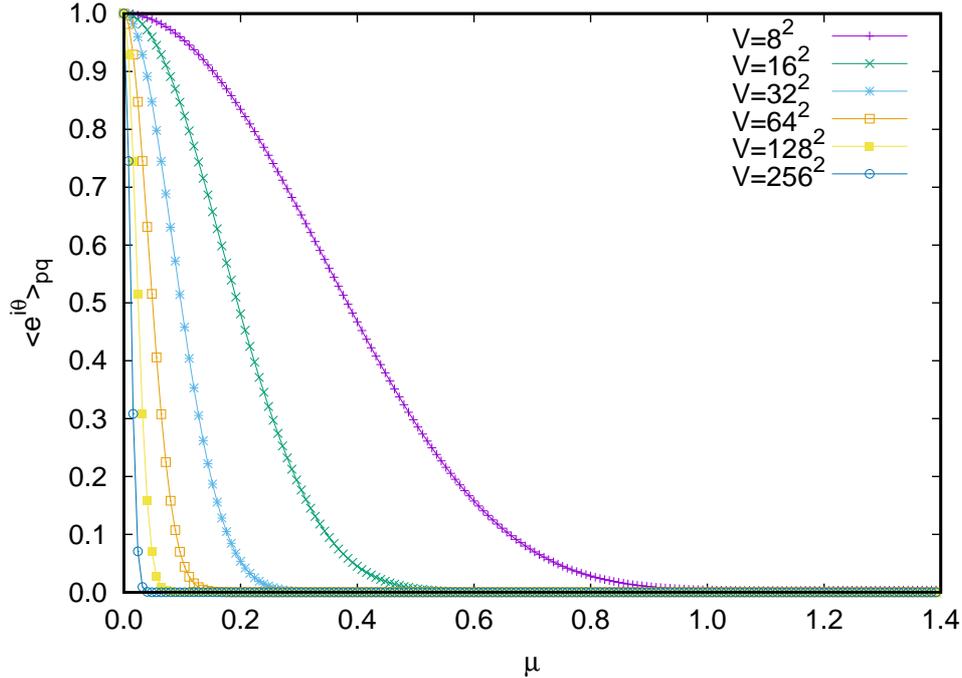}
  \caption{Average phase factor as a function of $\mu$. The parameters are $m^{2} = 0.01$, $\lambda=1$, $K=D=64$
    and $V=8^2,16^2, \ldots, 256^2$.
    The sign problem becomes severe for larger $\mu$ and $V$.}
  \label{fig:average_phase}
\end{figure}
Figure~\ref{fig:average_phase} shows the average phase factor evaluated by the TRG for various $\mu$ and $V$.
We use $m^2=0.01$ and $\lambda=1$ which are the same parameters as \cite{Orasch:2017niz}.
As clearly seen, the average phase factor decreases as $\mu$ increases for fixed space-time volume $V$
while it also decreases as $V$ increases for fixed $\mu$.
We thus confirm that, in the  zero temperature and large spacial volume limits,
severe sign problems happen even for small values of $\mu$.

\subsection{Silver Blaze phenomenon}
\label{sec:sb}

In the thermodynamic limit, bulk observables are independent of $\mu$ below a critical $\mu_c$
as well as finite density QCD.
This is called the Silver Blaze phenomenon which is a direct outcome of an imaginary part of the action.
Although the computational cost of the Monte Carlo method has a large volume dependence,
the TRG is suitable for observing the Silver Blaze phenomenon clearly since
its cost scales with the logarithm of the lattice volume and the thermodynamic limit can be easily taken.

\begin{figure}[htbp]
  \centering
  \includegraphics[width=0.8\hsize]{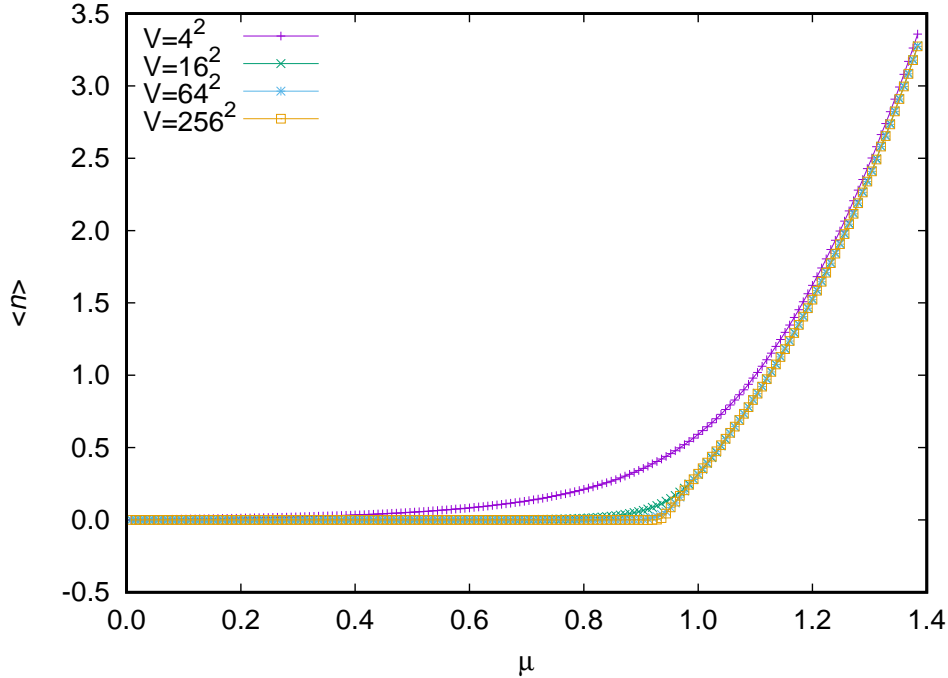}
  \caption{$\left< n \right>$ as a function of $\mu$.
    The lattice volume is varied from $4^2$ to $256^2$.
    The other parameters ($m$, $\lambda$, $K$ and $D$) are the same as those of fig.~\ref{fig:average_phase}.}
  \label{fig:number_density}
\end{figure}
Figure~\ref{fig:number_density} shows the $\mu$-dependence of particle number density,
\begin{align}
  \left< n \right>=\frac{1}{V}
  \frac{\partial \ln Z}{\partial\mu}.
\end{align}
The differentiation with respect to $\mu$ in the above equation is estimated by numerical differentiation.
The Silver Blaze phenomenon is clearly observed for large volumes.
The density does not depend on $\mu$ for
small $\mu$ region, and it begins to increase at $\mu\approx0.94$.
In particular, the cusp structure around $\mu\approx0.94$ tends to be sharper for larger volumes.

\begin{figure}[htbp]
  \centering
  \includegraphics[width=0.8\hsize]{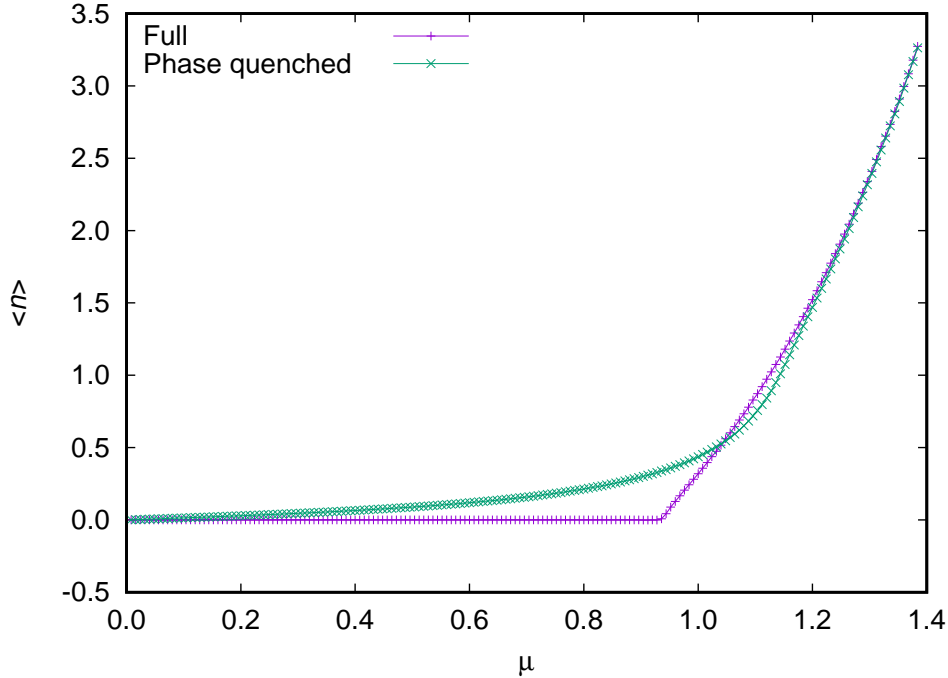}
  \caption{Comparison of the number density between the full and the phase quenched theories
    at $m^{2} = 0.01$, $\lambda=1$, $K=D=64$ on $V=1024^2$.
    The full theory clearly shows the Silver Blaze phenomenon unlike the phase quenched case.}
  \label{fig:comparison_phase_quench}
\end{figure}
\begin{figure}[htbp]
  \centering
  \includegraphics[width=0.8\hsize]{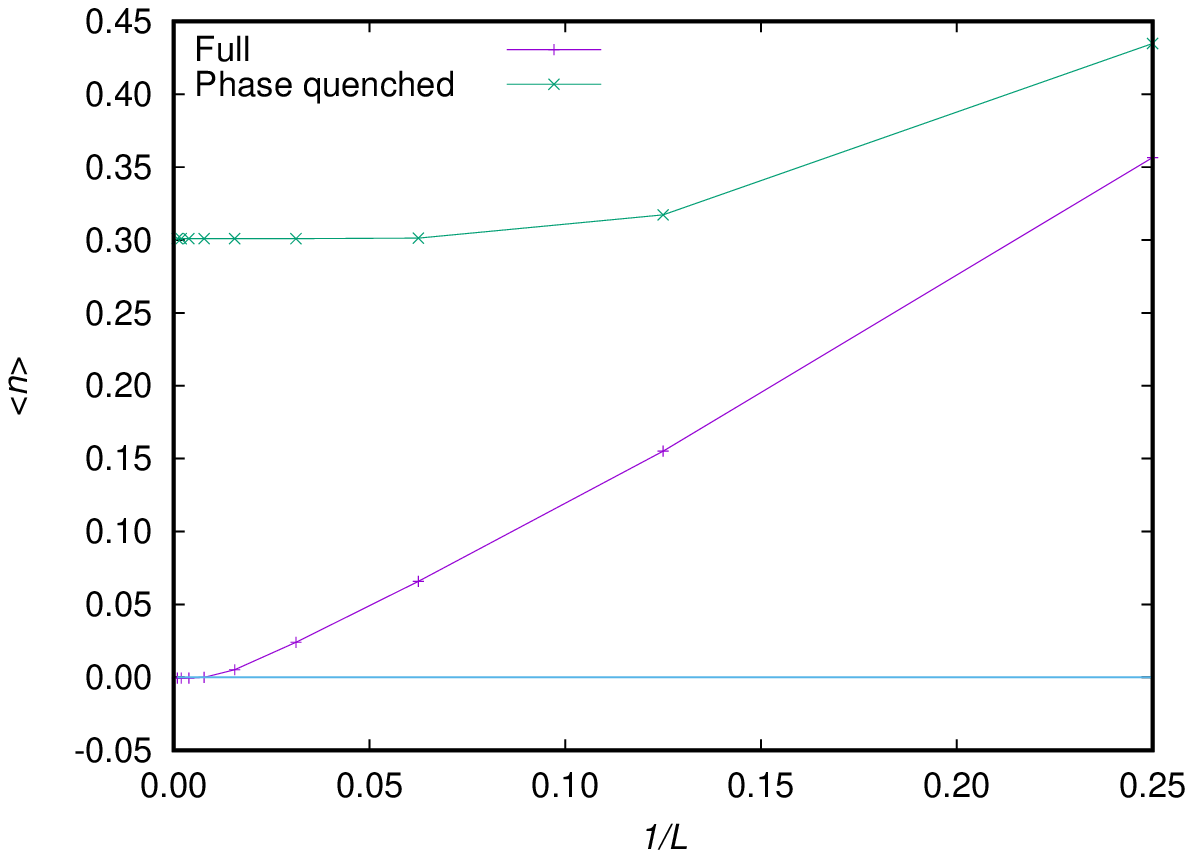}
  \caption{$\left< n \right>$ as a function of $1/L$
    at $\mu=0.904$, $m^{2} = 0.01$, $\lambda=1$, $K=D=64$.
    The density for the phase quenched case converges to a  non-zero value
    while that for the full theory converges to zero in the thermodynamic limit.}
  \label{fig:volume_dependence}
\end{figure}

In fig.~\ref{fig:comparison_phase_quench},  we compare the result of the number density
to that of the phase quenched model on $V=1024^2$.
By contrast to the full theory, the phase quenched model exhibits the continuous behavior,
and one can confirm that the $\mu$ independence of the result is a direct consequence of the imaginary part of the action.
To see the difference in more detail, the volume dependence of the result at $\mu=0.904$
is shown in fig.~\ref{fig:volume_dependence}.
In the infinite volume limit, although the result in the full theory converges to zero,  that in the
phase quenched model converges to a non-zero value.

\begin{figure}[htbp]
  \centering
  \includegraphics[width=0.8\hsize]{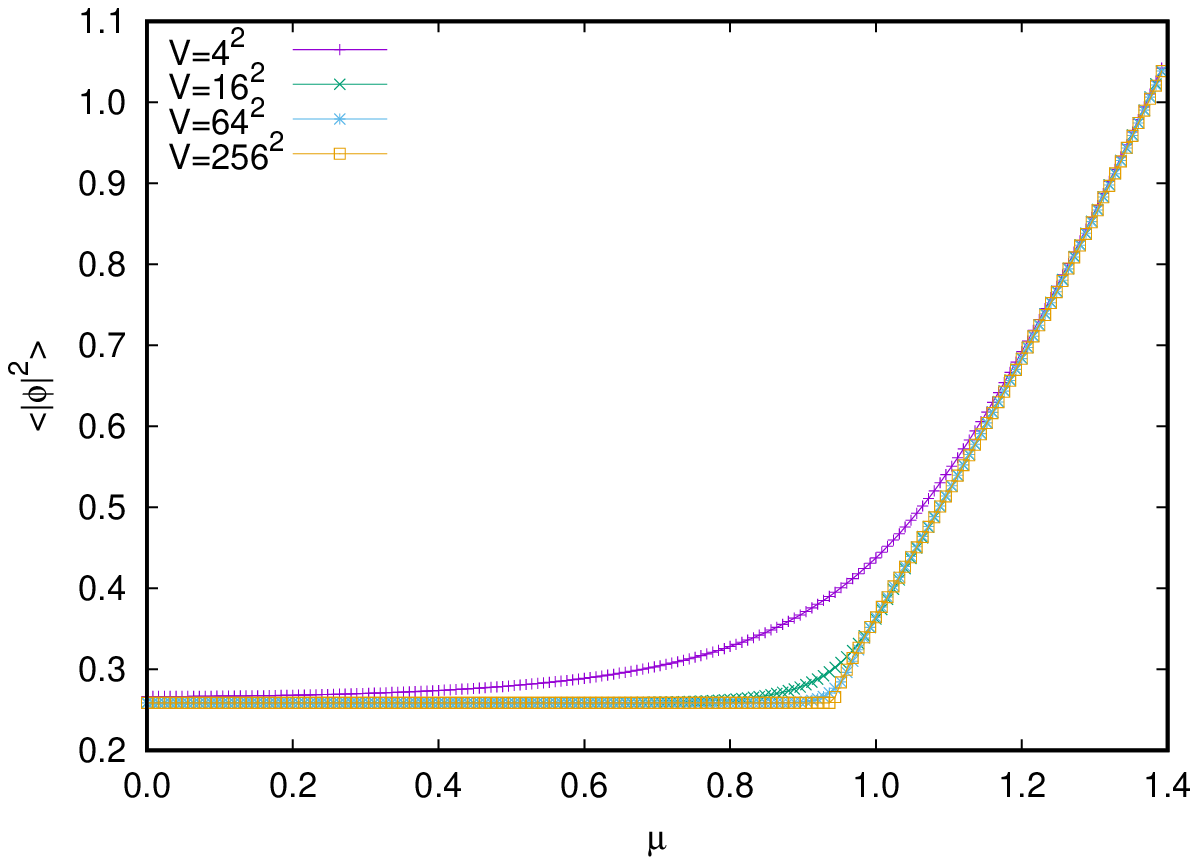}
  \caption{$\langle \left| \phi \right|^{2} \rangle$ as a function of $\mu$.
    The other parameters ($m$, $\lambda$, $K$, $D$ and $V$) are the same as those of fig.~\ref{fig:number_density}.
  }
  \label{fig:phi_squared}
\end{figure}
Figure~\ref{fig:phi_squared} shows $\langle \left| \phi \right|^{2} \rangle$ as a function of $\mu$ for
the same parameters as  those of fig.~\ref{fig:number_density},
which is evaluated
by the TRG with an impurity tensor~\cite{Kadoh:2018tis}.
As in the case of the density, the result is independent of $\mu$
for $\mu\lesssim0.94$ and a sharp rise is seen around $\mu\approx0.94$.

\begin{figure}[htbp]
  \centering
  \includegraphics[width=0.8\hsize]{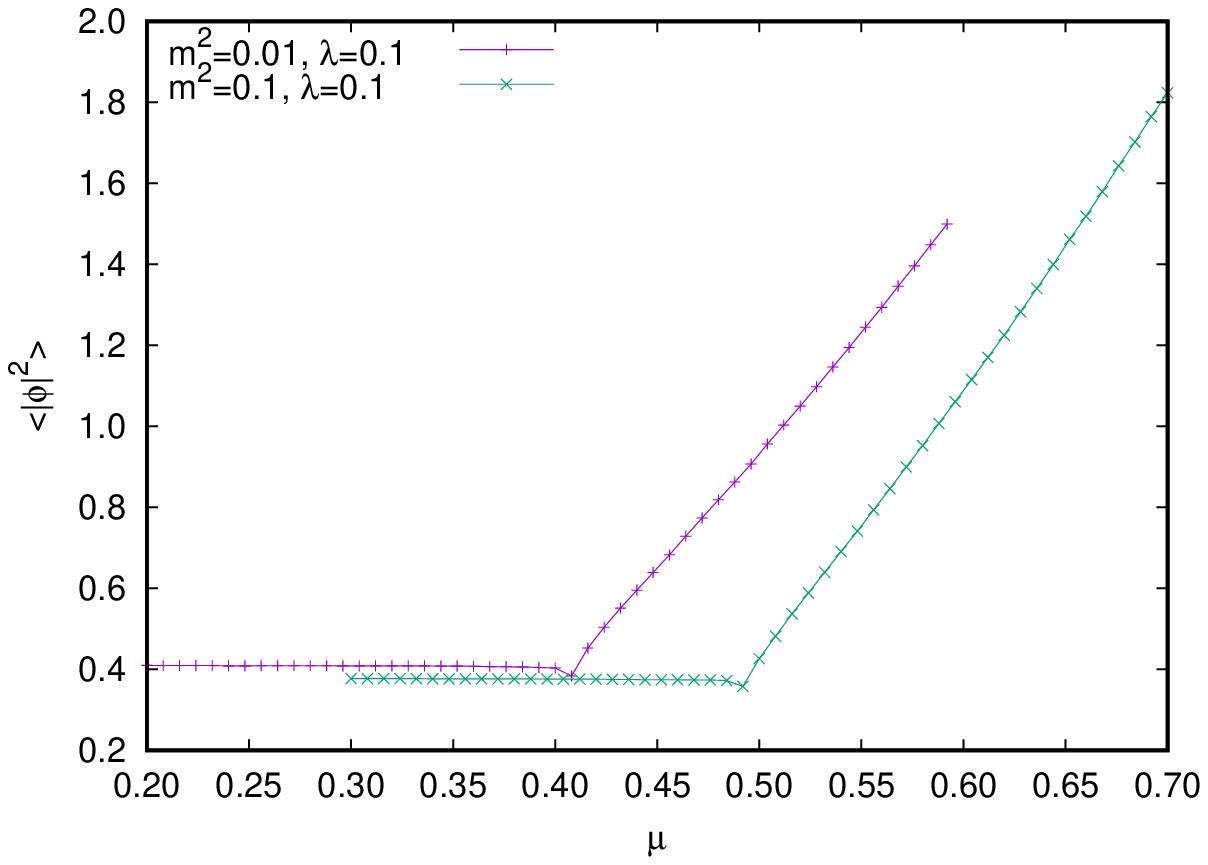}
  \caption{
    $\langle \left| \phi \right|^{2} \rangle$ as a function of $\mu$ at $K=64$ and $D=64$ on $V=1024^2$.
    The Silver Blaze phenomenon is observed irrespective of the values of the physical parameters.}
  \label{fig:difference_parameters}
\end{figure}
In order to study the stability of the Silver Blaze phenomenon
against changing the physical parameters ($m$ and $\lambda$),
we also compute the particle number density for $(m^{2},\lambda)=(0.01,0.1)$ and $(0.1,0.1)$ as shown in
fig.~\ref{fig:difference_parameters}.
Note that, for smaller $m$ or $\lambda$, the exponential damping in the Boltzmann weight is weaker.
Even for such cases, the Silver Blaze phenomenon is clearly observed.

\subsection{Comparison with another tensor network representation}

We have represented the partition function as a TN using the Gauss-Hermite quadrature
for both the real and imaginary parts of each scalar field
but one may use another
representation, for instance, with a polar coordinate and the character expansion
given in Appendix \ref{sec:character_expansion}.
It is known that  the partition function in the case does not have an imaginary part.
This formulation is also useful for the TN method.
The Gaussian quadrature is needed only for  the radial variable because the angular variable is transcribed
into a tensor index with the character expansion.
Thus, the cost of making the initial tensor is basically cheaper than making eq.~(\ref{eq:tensor}).
See Appendix \ref{sec:character_expansion} for the details.

\begin{figure}[htbp]
  \centering
  \includegraphics[width=0.8\hsize]{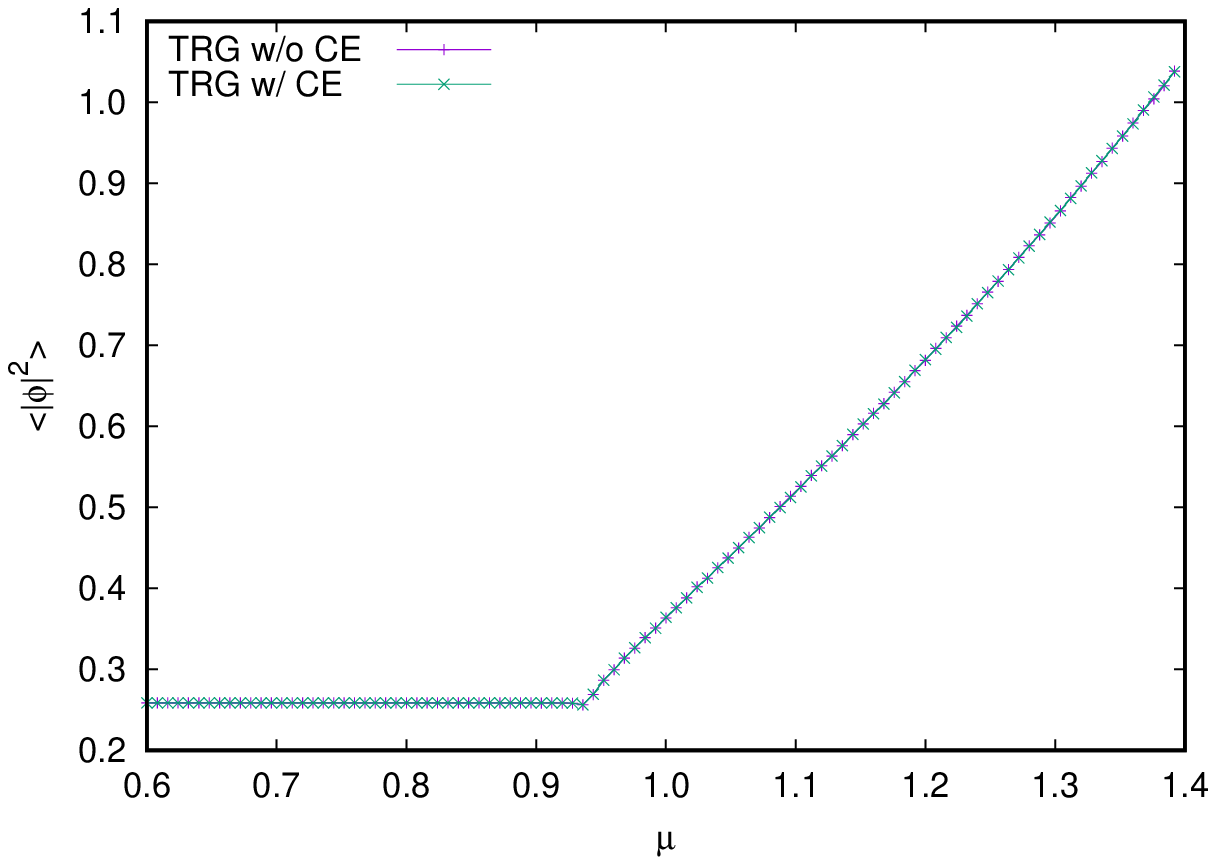}
  \caption{Comparison of
    $\langle \left| \phi \right|^{2} \rangle$ obtained by two different TN formulations with $D=64$ on $V=1024^2$.
    We use $K=64$ for the first representation without the character expansion
    and  $N_{CE}=128$ and $K=256$ for the second one with  the character expansion.
    See the appendix for the details.
  }
  \label{fig:comparison}
\end{figure}
In figure~\ref{fig:comparison}, two representations are compared by showing $\langle|\phi^2|\rangle$  against $\mu$.
As a result they agree with each other well and
it is hard to see the difference between them at this resolution.
Thus we can conclude that choices of TN representation are basically irrelevant to our conclusion.

\section{Summary}
\label{sec:summary}

In this paper we have derived a TN representation for the complex scalar field theory
discretizing the continuous scalar fields with  the Gauss-Hermite quadrature rule.
Using the TRG procedure for the TN representation of partition function,
the average phase factor, the particle number density and $\langle |\phi|^2\rangle$ were evaluated.

As a result, the Silver Blaze phenomenon is clearly observed for
the extremely large volume $V=1024^2$ which is essentially
in the zero temperature and the large spacial volume limits.
We also examine another TN representation using the character expansion.
Then, our numerical results of two representations
do not have a visible difference, and the conclusion does not change for the other TN representation.
Thus we confirm that the TN method is effective
for a quantum field theory with the severe sign problem.

\section*{Acknowledgments}

We would like to thank Yoshimasa Hidaka and Akira Onishi for their helpful comments.
This work was supported by
the Ministry of Education, Culture, Sports, Science and Technology (MEXT) as ``Exploratory Challenge on Post-K computer'' (Frontiers of Basic Science: Challenging the Limits),
JSPS KAKENHI Grant Number JP17K05411, JP18J10663 and JP19K03853.

\appendix

\section{Tensor network representation with a character expansion}
\label{sec:character_expansion}
In this appendix, an alternative TN representation of partition function is
derived by a character expansion (See \cite{Endres:2006xu})\footnote{
  If the Taylor expansion of the hopping term is used instead of the character expansion,
  another dual formulation is obtained\cite{Orasch:2017niz}.
}.
To this end, the polar coordinate $\phi_n=r_n e^{i\theta_n}$ is helpful in using
the character expansion of  $e^{x \cos z}$:
\begin{align}
  \label{eq:ce}
  e^{x \cos z} = \sum_{p = -\infty}^{\infty} I_{p}\left( x \right) e^{ipz} \qquad \quad  \text{for } x \in \mathbb{R},\ z \in \mathbb{C},
\end{align}
where $I_{p}$ is the $p$-th modified Bessel function of the first kind.
In the polar coordinate, the lattice action eq.~(\ref{eq:laction}) is written as
\begin{align}
  S=\sum_{n\in \Gamma} \left[
  \left( 4 + m^{2} \right) r_{n}^{2}
  + \lambda  r_{n}^{4}
  - 2\sum_{\nu=1}^{2} \cos (\theta_{n+\hat{\nu}} - \theta_n -i \mu \delta_{\nu 2})  r_{n}r_{n+\hat{\nu}}
  \right].
\end{align}
Using the above formulas, a dual formulation of the partition function is obtained as
\begin{eqnarray}
  &&        Z = \left( \prod_{n\in\Gamma} \sum_{p_{n},q_{n}=-\infty}^{\infty} \int_{0}^{\infty} \mathrm{d}r_{n} r_n \,
     e^{ - \left( 4 + m^{2} \right) r_{n}^{2} - \lambda  r_{n}^{4} } \right) \nonumber \\
  &&      \qquad \times
     \prod_{n\in\Gamma}\,
     I_{p_n}(2 r_n r_{n+\hat1}) I_{q_n}(2 r_n r_{n+\hat2})
     \, \delta_{(p_{n}+q_{n}- p_{n-\hat{1}}-q_{n-\hat{2}}),0} \, e^{\mu q_{n}}.
     \label{eq:zce}
\end{eqnarray}
Integrating the angular variables turns out to be  constraints for $p$ and $q$ variables with Kronecker's delta.
Note that all entries of (\ref{eq:zce}) are real and non-negative.

To define a finite dimensional tensor, we truncate the summation for $p_n, q_n$
and discretize the radial variable $r_n$ with Gauss-Hermite quadrature.
In this case, since $r_n \in \left[ 0, \infty \right)$,
we use the $2K$-point Gauss--Hermite quadrature with only the positive $K$ nodes~\footnote{
  One can of course use other quadrature rules such as the Gauss--Legendre or the Gauss--Laguerre.
  With sufficiently large number of Gaussian nodes, the detail of the quadrature rule does not matter to the accuracy.
}.
Then the discrete version of \eqref{eq:zce} is given by
\begin{eqnarray}
  &&  Z\left( N_{\mathrm{CE}}, K \right)
     = \left( \prod_{n\in\Gamma} \sum_{p_{n},q_{n}=-N_{\mathrm{CE}}}^{N_{\mathrm{CE}}} \sum_{\alpha_{n}=1}^{K} \right) \prod_{n\in\Gamma}
     2 \pi y_{\alpha_{n}} w_{\alpha_{n}} e^{y_{\alpha}^{2}}
     \nonumber \\
  && \qquad \quad \times \, h_{p_{n}}( y_{\alpha_{n}}, y_{\alpha_{n+\hat{1}}} ) h_{q_{n}}( y_{\alpha_{n}}, y_{\alpha_{n+\hat{2}}} )
     \,
     \delta_{(p_{n}+q_{n}-p_{n-\hat{1}}-q_{n-\hat{2}}), 0} \,  e^{\mu q_{n}},
     \label{eq:dpfunction_ce}
\end{eqnarray}
where
\begin{align}
  h_{p} \left( r, s \right)
  = e^{- \left(1+ \frac{m^{2}}{4} \right) \left( r^{2} + s^{2} \right) - \frac{\lambda}{4} \left( r^{4}+s^{4} \right)} I_{p}\left( 2rs \right),
\end{align}
$y_\alpha$ is the $\alpha$th positive root of
$2K$ Gauss--Hermite polynomial,  and $w_\alpha$ is the corresponding weight given by
$w_\alpha \equiv 2^{2K-1} (2K)!\sqrt{\pi}/((2K)^2H_{2K-1}(y_\alpha)^2)$.
We have restricted the range of the summation of  the character expansion to $\left[ -N_{\mathrm{CE}}, N_{\mathrm{CE}} \right]$.
The local Boltzmann factor $h_p(y_{\alpha},y_{\beta})$ with fixed $p$ is now regarded as a $K\times K$ matrix
to which the SVD can be applied:
\begin{align}
  \label{eq:hSVD}
  h_{p}\left( y_\alpha, y_\beta \right)
  = \sum_{x=1}^{K} U^{[p]}_{\alpha x} \sigma^{[p]}_{x} V^{[p]\dagger}_{x\beta}.
\end{align}
Plugging eq.~(\ref{eq:hSVD}) into eq.~(\ref{eq:dpfunction_ce}) leads to
a TN representation of $Z\left( N_{\mathrm{CE}}, K \right)$:
\begin{align}
  \label{eq:ZCE}
  Z\left( N_{\mathrm{CE}}, K \right)
  = \left( \prod_{n\in\Gamma} \sum_{p_{n},q_{n}=-N_{\mathrm{CE}}}^{N_{\mathrm{CE}}} \sum_{x_{n}, t_{n}=1}^{K} \right) \prod_{n\in\Gamma}
  \begin{aligned}[t]
    & \tilde T^{\, p_{n} q_{n} p_{n-\hat{1}} q_{n-\hat{2}}}_{\, x_{n} t_{n} x_{n-\hat{1}} t_{n-\hat{2}}},
  \end{aligned}
\end{align}
where
\begin{align}
  \tilde T_{\, ijkl}^{\, abcd}
  =
  2\pi\sqrt{\sigma^{[a]}_{i} \sigma^{[b]}_{j} \sigma^{[c]}_{k} \sigma^{[d]}_{l}} \,  e^{\mu b} \, \delta_{a+b,c+d}
  \, \sum_{\alpha=1}^{K} y_{\alpha} w_{\alpha} e^{y_{\alpha}^{2}} U^{[a]}_{\alpha i} U^{[b]}_{\alpha j} V^{[c]\dagger}_{k \alpha} V^{[d]\dagger}_{l \alpha}.
  \label{tensor_ce}
\end{align}
Note that $(x_n,p_n), (t_n,q_n),(x_{n-\hat 1}, p_{n-\hat 1}),(t_{n-\hat 2},q_{n-\hat 2})$ may be
interpreted as four index pairs defined on four different links stemmed from the site $n$. Thus $\tilde T$
may be interpreted as a rank-4 tensor whose bond dimension is $K\times(2N_{\rm CE}+1)$
since $x_n,t_n=1,2\ldots,K$ and $p_n,q_n=-N_{\rm CE}, \ldots, N_{\rm CE}$.

In an actual computation, the summations $\sum_{p_n=-N_{\rm CE}}^{N_{\rm CE}} \sum_{x_n=1}^K$ (and $\sum_{q_n=-N_{\rm CE}}^{N_{\rm CE}} \sum_{t_n=1}^K$)
in eq.~(\ref{eq:ZCE}) are reduced
by including $D$ largest singular values $\sigma_x^{[p]}$ of eq.~(\ref{tensor_ce}) into the computation.
Let us arrange $\sigma_x^{[p]}$ in the descending order for all $x$ and $p$ and suppose that the $n$th largest singular value
is  $\sigma_{x'}^{[p']}$. Then a one-to-one mapping $f$ between $n$ and $(x',p')$ can be given, that is, $n=f(x',p')$.
\footnote{For degenerate singular values, one may give the mapping in arbitrary order. }
Using this mapping, the combined index $X_n$ and $T_n$ are given by
\begin{align}
  X_n=f(x_n,p_n), \quad T_n =f(t_n,q_n).
\end{align}
Once $f$ is given,  $X_n$ is uniquely given  for $x_n,p_n$ and vice versa.
Then the tensor is represented as
\begin{align}
  T^{({\rm CE})}_{X T X^\prime T^\prime } \equiv
  \tilde T^{p q p^\prime q^\prime}_{x t x^\prime t^\prime},
\end{align}
with $(x,p)=f^{-1}(X), (t,q)=f^{-1}(T)$ and the same identifications for $X^\prime,T^\prime$.
Then truncated version of the discretized partition partition function is given by
\begin{align}
  \label{ZCE_final}
  Z\left( N_{\mathrm{CE}}, K \right)
  \approx \left( \prod_{n\in\Gamma} \sum_{X_{n}, T_{n}=1}^{D} \right) \prod_{n\in\Gamma}
  \begin{aligned}[t]
    & T^{({\rm CE})}_{X_n T_n X_{n-\hat1} T_{n-\hat2}}.
  \end{aligned}
\end{align}
For a given $D$ and $K$, the mapping $f$  is uniquely determined up to $D$ (except for an issue of degenerate singular values)
because it is stable in practice when $N_{\rm CE}$ is varied in a range of sufficiently large values.
So $T^{({\rm CE})}$ does not depend on $N_{\rm CE}$, but only on $K$.
In this sense, the $N_{\rm CE}$-dependence of $Z\left( N_{\mathrm{CE}}, K \right)$ is small and is not observed
as long as it is approximated by r.h.s. of eq.~(\ref{ZCE_final}).

\end{document}